\documentclass{aastex63}
\usepackage{lineno}
\usepackage{hyperref}
\usepackage[utf8]{inputenc}
\usepackage{amsmath}
\usepackage{amsfonts}
\usepackage{enumerate}
\usepackage{ytableau}
\usepackage{graphicx}
\usepackage{indentfirst}
\usepackage[english]{babel}
\usepackage{overpic}
\usepackage[normal,footnotesize]{caption2}
\def\baselinestretch{1.25}
\usepackage{appendix}
\usepackage{amssymb}
\usepackage{amsthm}
\usepackage{epsfig,graphicx,epstopdf}

\def\gray{$\gamma$-ray\ }
\def\grays{$\gamma$-rays\ }
\def\fermi{{\it Fermi }}

\received{***}
\revised{***}
\accepted{***}
\submitjournal{ApJL}

\shorttitle{LHAASO J0341+5258}
\shortauthors{LHAASO Collaboration}

\begin{document}
\title{ Discovery of a new $\gamma$-ray source LHAASO J0341+5258 with emission up to 200TeV}

\author{Zhen Cao}
\affiliation{Key Laboratory of Particle Astrophyics \& Experimental Physics Division \& Computing Center, Institute of High Energy Physics, Chinese Academy of Sciences, 100049 Beijing, China}
\affiliation{University of Chinese Academy of Sciences, 100049 Beijing, China}
\affiliation{TIANFU Cosmic Ray Research Center, Chengdu, Sichuan,  China}

\author{F. Aharonian}
\affiliation{Dublin Institute for Advanced Studies, 31 Fitzwilliam Place, 2 Dublin, Ireland }
\affiliation{Max-Planck-Institut for Nuclear Physics, P.O. Box 103980, 69029  Heidelberg, Germany}

\author{Q. An}
\affiliation{State Key Laboratory of Particle Detection and Electronics, China}
\affiliation{University of Science and Technology of China, 230026 Hefei, Anhui, China}

\author{Axikegu}
\affiliation{School of Physical Science and Technology \&  School of Information Science and Technology, Southwest Jiaotong University, 610031 Chengdu, Sichuan, China}

\author{L.X. Bai}
\affiliation{College of Physics, Sichuan University, 610065 Chengdu, Sichuan, China}

\author{Y.X. Bai}
\affiliation{Key Laboratory of Particle Astrophyics \& Experimental Physics Division \& Computing Center, Institute of High Energy Physics, Chinese Academy of Sciences, 100049 Beijing, China}
\affiliation{TIANFU Cosmic Ray Research Center, Chengdu, Sichuan,  China}

\author{Y.W. Bao}
\affiliation{School of Astronomy and Space Science, Nanjing University, 210023 Nanjing, Jiangsu, China}

\author{D. Bastieri}
\affiliation{Center for Astrophysics, Guangzhou University, 510006 Guangzhou, Guangdong, China}

\author{X.J. Bi}
\affiliation{Key Laboratory of Particle Astrophyics \& Experimental Physics Division \& Computing Center, Institute of High Energy Physics, Chinese Academy of Sciences, 100049 Beijing, China}
\affiliation{University of Chinese Academy of Sciences, 100049 Beijing, China}
\affiliation{TIANFU Cosmic Ray Research Center, Chengdu, Sichuan,  China}

\author{Y.J. Bi}
\affiliation{Key Laboratory of Particle Astrophyics \& Experimental Physics Division \& Computing Center, Institute of High Energy Physics, Chinese Academy of Sciences, 100049 Beijing, China}
\affiliation{TIANFU Cosmic Ray Research Center, Chengdu, Sichuan,  China}

% ...

\begin{abstract}
We report the discovery of a new unidentified extended $\gamma$-ray source in the Galactic plane named LHAASO J0341+5258 with a pre-trial significance of 8.2 standard deviations above 25 TeV. The best fit position is R.A.$=55.34^{\circ}\pm0.11^{\circ}$ and Dec$=52.97^{\circ}\pm0.07^{\circ}$. The angular size of LHAASO J0341+5258 is $0.29^\circ \pm 0.06^\circ_{stat} \pm0.02^\circ_{sys}$. The flux above 25 TeV is about $20\%$ of the flux of Crab Nebula.  Although a power-law fit of the spectrum from 10 TeV to 200 TeV with the photon index $\alpha=2.98 \pm 0.19_{stat} \pm 0.02_{sys}$ is not excluded,  the LHAASO data together with the flux upper limit at 10  GeV set by the Fermi LAT observation, indicate 
a noticeable  steepening of an initially hard power-law spectrum 
spectrum  with a cutoff  at  $\approx 50$~TeV.  We briefly discuss the origin  of  UHE gamma-rays. The lack of an energetic pulsar and a young SNR inside or in the vicinity of   LHAASO J0341+5258 challenge, but do not exclude both the leptonic and hadronic scenarios of gamma-ray production. 

\end{abstract}


\section{Introduction} \label{sec:intro}
The distinct cosmic ray spectral feature around $10^{15}$ eV, the so-called “knee”, in the locally measured cosmic ray spectrum implies that the galaxy contains objects accelerating protons and nuclei to PeV energies. Identification of these accelerators or the so-called PeVatrons is a prime objective towards the understanding of the origin of galactic cosmic rays. Gamma rays and neutrinos, produced in the interaction of cosmic rays with the ambient medium within or around the accelerator, are key signatures of these cosmic ray factories.  

Over the last two decades, more than two hundred  very high energy (VHE, $>$0.1 TeV) \gray sources have been reported \footnote{\url{http://tevcat.uchicago.edu/}} due to the successful operation of imaging atmospheric Cherenkov telescopes (IACTs) and extensive air shower (EAS) arrays. Although there is much progress made in recent years, the origin of cosmic rays is still an open question. The spectrum of most of the reported gamma-ray sources can be well explained by the inverse Compton scattering of directly accelerated electrons. Further exploration in the TeV-PeV energy band is necessary to pin down which kind of source could explain the cosmic rays up to the knee and beyond.

The proposed possible candidates for galactic PeVatron include supernova remnants (SNRs)(for a review see e.g.\citet{2013APh....43...71A}), young massive star clusters \citep{2019NatAs...3..561A}, the super-massive black hole in the Galactic Center\citep{2016Natur.531..476H}, pulsars \citep{bednarek02}, etc.  Observations of the \gray sky above tens of TeV, especially in ultra-high energy band (UHE, $>$0.1 PeV), is crucial for searching  and identifying PeVatrons. The current generation of IACTs are most sensitive at TeV energies. However, the low flux  of photons  above 10 TeV requires large exposure which is limited by the duty-cycle and few degree field of view (FoV) of IACTs. 
On the other hand, EAS arrays, such as Tibet-As$\gamma$ and HAWC, have high duty-cycle and wide field of view, so that they can detect gamma-ray sources beyond energies of tens of TeV\citep{2019PhRvL.123e1101A,2020PhRvL.124b1102A}.

LHAASO is a large hybrid EAS array being constructed on the Haizi Mountain, Daocheng, Sichuan province, China. It is composed of three sub-arrays, including the one square-km array (KM2A), Water Cherenkov Detector Array (WCDA) and  the wide-field air Cherenkov/fluorescence telescopes (WFCTA) array \citep{article}. Being the most sensitive observatory above a few tens of TeV currently in operation, LHAASO can survey the sky in the declination band from -15$^{\circ}$  to 75$^{\circ}$  with full duty-cycle and has the potential to find new sources at UHE \citep{2019ChA&A..43..457C}. Even though only a half of KM2A has been operated since the end of 2019, the sensitivity is already better than what have been achieved by previous observations above tens of TeV. Benefiting from the excellent sensitivity, LHAASO has detected 12 UHE sources with the statistical significance greater than 7$\sigma$\citep{2021Natur.594...33C}. The detailed study of the performances of KM2A via the observation on the Crab Nebula is presented in \citet{2021ChPhC..45b5002A}. In this paper, we report on the detection and study of primary properties of a newly discovered source above tens of TeV. Possible counterparts in other wavelengths and different possible  production scenarios are discussed.

\section{The LHAASO-KM2A experiment}
The whole KM2A array consists of 5195 electromagnetic particle detectors (EDs) and 1188 muon detectors (MDs), which are spread over 1.3 km$^2$ area. The EDs and MDs are used to detect the number of second particles and record their arriving time. The ED is a scintillation detector covered by a 5-mm-thick lead plate, to absorb low-energy charged particles and to convert high energy $\gamma$-rays into electron-positron pairs. The MD is a water Cherenkov detector with ultra-pure water as detection medium.  The detector is covered by overburden soil with thickness of 2.5 m, which absorbs most of the secondary
electron/positrons and $\gamma$-rays. More details about the detectors are presented in \citet{article}.  About half of the detectors have been installed and started operation since the end of 2019.

Once more than 20 EDs fired within a time window of 400 ns, a shower will trigger the array, and all hits within $5\,{\mu}$s before and after the trigger time will be recorded.  An off-line method is developed to calibrate the time measurements of EDs. The charge measurements of EDs are calibrated by selecting single minimum ionization particle (MIP) events in the EAS. More details about ED calibration can be found elsewhere \citep{LV201822}. The calibration of time  and charge responses for MDs is similar to the calibration of EDs. With the calibration of each detector, the ADC counts of EDs and MDs are converted into number of particles. The status of each detector is monitored in real time, and only detectors under normal conditions are used for reconstruction. 
 
 Both measured and simulated events are processed through the same reconstruction pipeline. The direction of showers is reconstructed by fitting the relative arriving time of the ED hits. The angular resolution (denoted as $\phi_{68}$, containing $68\%$ 

1 warning
 of the events) is ${0.5^{\circ}} - {0.8^{\circ}}$ at 20 TeV and $0.24^{\circ} - 0.3^{\circ}$ at 100 TeV for \gray showers at different zenith angles. The parameter $\rho_{50}$, defined as the particle density at 50 m from the shower axis obtained by fitting the Nishimura-Kamata-Greisen (NKG) function to the shower hits, is used as energy estimator\citep{2021ChPhC..45b5002A}. The energy resolution is about $24\%$ at 20 TeV and $13\%$ at 100 TeV for showers with zenith angle less than $20^{\circ}$. Considering there are more muons and less electrons in hadronic showers than electromagnetic showers of the same energy, the ratio between muons and electrons is used to discriminate electromagnetic showers from hadronic showers. The survival ratio for protons and nucleus is about $1\times10^{-3}$ at 40 TeV with $48\%$ of gamma-rays survived. The rejection power will be better at higher energies. The data reconstruction, calibration and selection are described detailed  in \citet{2021ChPhC..45b5002A}.
 
\section{Analysis and Results}
The data used in this analysis were collected from December 2019 to November 2020 by the half of KM2A. The total effective observation time is 308.33 days after the data quality check, and then the same event reconstruction and selection are used as described in the performance paper \citep{2021ChPhC..45b5002A}. The dedicated observation time on LHAASO J0341+5258 is 2368.7 hours.

Considering the energy resolution and statistics, one decade of energy is divided into 5 bins with a bin width of $log_{10}E = 0.2$. The sky in celestial coordinates (right ascension and declination) is divided into grids with size of $0.1^{\circ}\times0.1^{\circ}$  and filled with detected events according to their reconstructed arrival directions for each energy bin. The ``direct integration method'' \citep{2004ApJ...603..355F} is adopted to estimate the number of cosmic ray background events. This method uses events with the same direction in
horizontal coordinates but at different arrival time to estimate
the background. An integration of 24 hours of off-source
data is used to estimate background in this work. There are some local effects, such as the galactic diffuse gamma emission, which may also influence the results. The ratio of the total events to the total background events at a distance between $1.5^{\circ}$ and $3.5^{\circ}$ from the center of the source is used to remove this effect. The test statistic used to evaluate the significance of the test is TS $= 2 log (\lambda)$, where $\lambda={\mathcal{L}_{s+b}}/{\mathcal{L}_{b}}$. ${\mathcal{L}_{s+b}}$ is the maximum likelihood value for source plus background hypothesis, while $\mathcal{L}_{b}$ is the  background-only hypothesis. According to Wilks' Theorem \citep{wilks1938}, TS follows chi-square distribution with the number of degrees of freedom equal to the difference of number of free parameters between the hypotheses. In this work, a two-dimension Gaussian model with sigma fixed at $\phi_{68}/1.51$ is used, and the only free parameter is the total number of events. Thus we can take $\pm\sqrt{\rm TS}$ as the significance of observed results.  

\begin{figure*}
\plotone{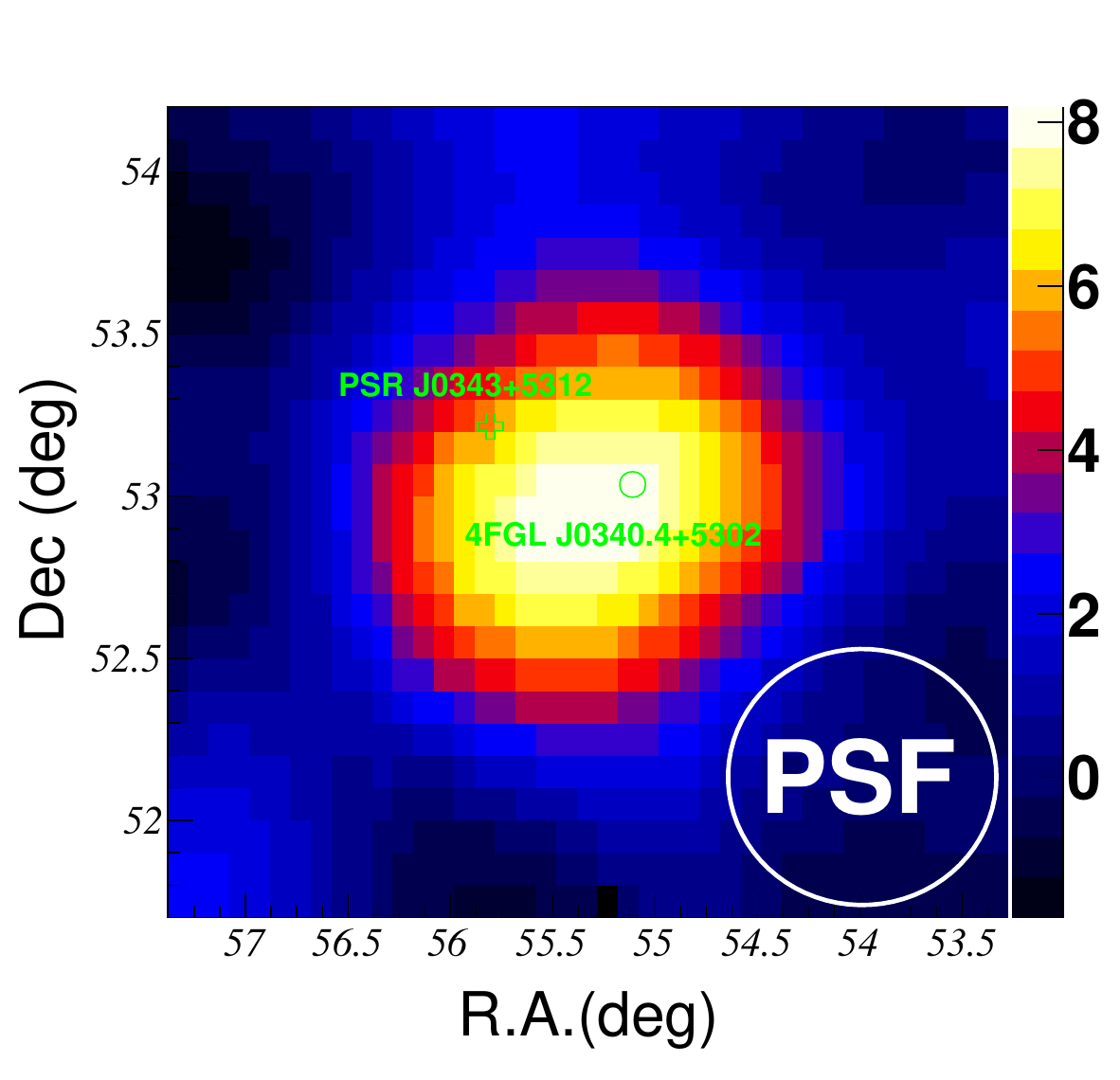}
\caption{The significance map of LHAASO J0341+5258 above 25 TeV. The green circle marks the position of 4FGL J0340.4+5302, and the blue cross marks the position of the pulsar PSR J0343+5312. The white circle at the bottom-right corner shows the size of the PSF ($68\%$ containment). The PSF value is $0.40^{\circ}$ above 25 TeV.}
\label{Fig.1}
\end{figure*}

 The significance map for energies above 25 TeV is shown in Figure \ref{Fig.1}, and the highest pre-trial significance value is $8.2\sigma$. Probabilities must be scaled by a trial factor accounting for the number of attempts to find a source, here conservatively assumed to be equal to the number of bins in the sky map (3600$\times$900). The significance after trials is about 6.0 $\sigma$. It should be noted that the trials is over-estimated here; since the smoothing radius is larger than the bin width, the significance in adjacent bins are correlated.

The method to determine the position and extension of the source is similar to the above description.  However the position and extension of the source are left as free parameters. The intrinsic extension is determined to be $\sigma_{\rm ext}=0.29^{\circ}\pm{0.06}^{\circ}$. The best fit position is R.A.$=55.34^{\circ}\pm0.11^{\circ}$ and Dec$=52.97^{\circ}\pm0.07^{\circ}$. The potential inconsistency between simulation  and data on the point spread function is the main systematic error, which has an impact of 0.02$^{\circ}$ on extension. To study the significance of extension of the source, we compared the $\Delta{\rm TS}$ between extensive and point source assumptions, which results in a $\Delta{\rm TS}$ of 13.3. The extensive source assumption is favored at $3.6\sigma$.

\begin{figure}
\plotone{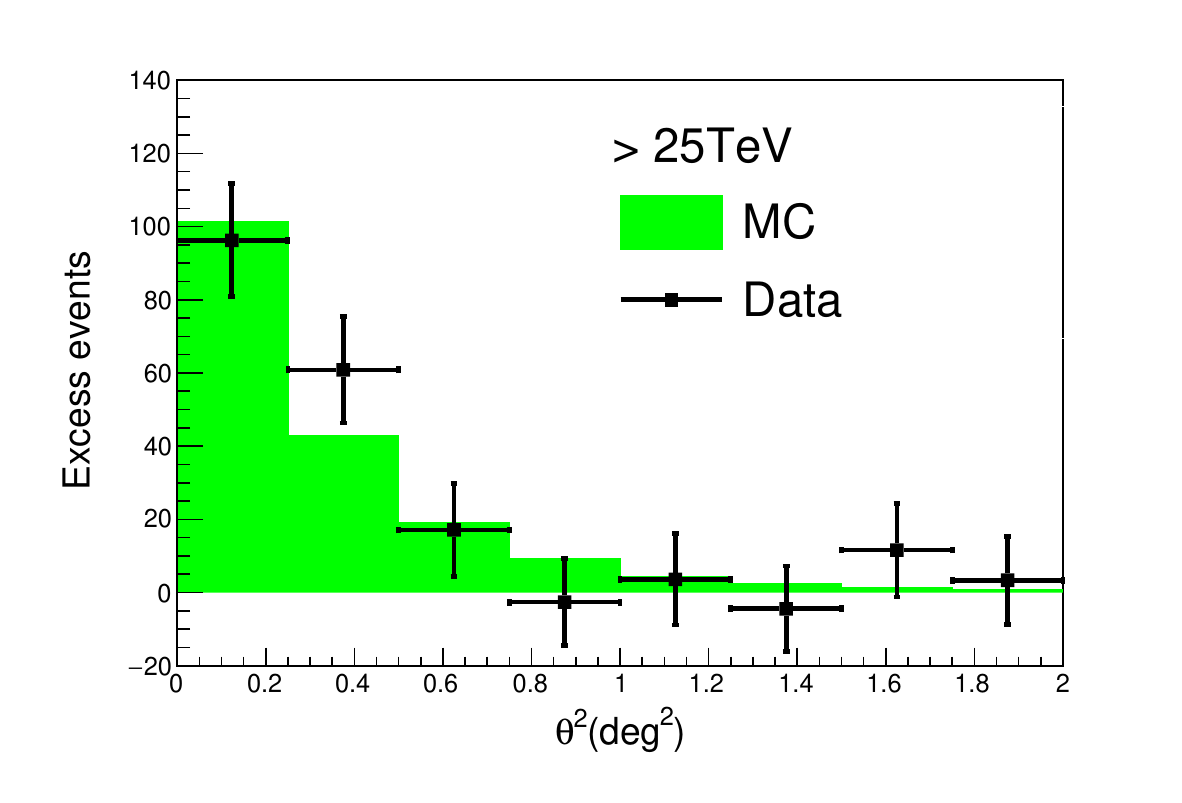}
\caption{Distribution of events as a function of the square of the angular distance to the LHAASO J0341+5258 direction   for both experimental data (black points) and MC simulation (green area). Events with energy above 25 TeV are used here. }
\label{Fig.2}
\end{figure}

The intrinsic extension of J0341+5258 is checked  by comparing the distribution of $\theta^{2}$ for signals between experiment and simulation as shown in Figure \ref{Fig.2}, where $\theta$ is the angular distance of each event to the position of J0341+5258. A set of $\gamma$-rays are generated taking into account the spectral energy distribution (SED), the intrinsic source extension and the detector PSF. The good agreement between simulation and experimental results with $\chi^2/ndf=7.4/8$ demonstrates the 
correct understanding of the extension of source.

\begin{figure}
\centering
\plotone{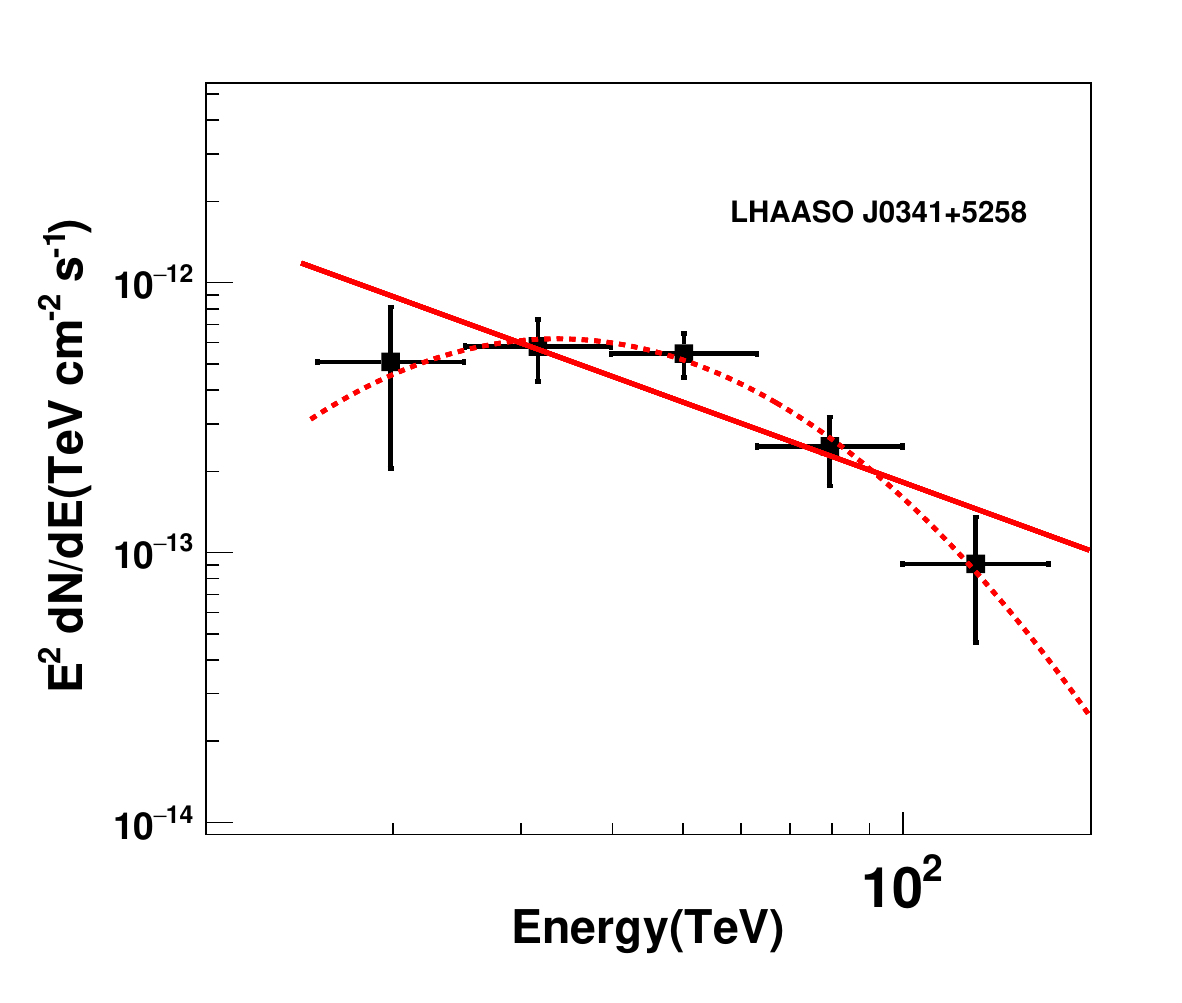}
\caption{The energy spectrum of LHAASO J0341+5258. Only statistic errors are shown here. The solid line is the best fit result assuming a single power-law function and the dotted line is the result from log-parabola function.}
\label{Fig.3}
\end{figure}

We estimate the spectrum of LHAASO J0341+5258 using a forward-folding method, described in \citet{2021ChPhC..45b5002A}. The total number of events in each energy bin is generated by simulating the detector's response to each event and assuming a power-law spectrum {\bf $f(E)=J\cdot(E/E_{0})^{-\alpha}$. The reference energy $E_{0}$ is chosen to be 40~TeV here.} The best-fit values of J and $\alpha$ are obtained by minimizing $\chi^2$ function. The Spectral energy distribution($E^2 dN/dE$) is shown in Figure \ref{Fig.3}. The differential flux (TeV$^{-1}$cm$^{-2}$s$^{-1}$) in the energy range from 10 TeV to 200 TeV is:
\begin{equation}
\frac{dN}{dE dA dt}=(2.8\pm0.4_{sta}\pm0.2_{sys})\times10^{-16}(\frac{E}{40~\rm TeV})^{-2.98\pm0.19_{stat}\pm0.02_{sys}} 
\end{equation}

The $\chi^2/ndf$ for the fit is 6.58/3. The integral flux above 25 TeV is $1.43\times10^{-14}(~\rm cm^{-2}s^{-1})$, corresponding to about $20\%$  of the flux of Crab Nebula. We should note that in general the atmospheric density profile always deviates from the atmosphere model used in simulations, which is the main systematic error affecting the SED. The total systematic uncertainty is estimated to be $7\%$ for the flux and 0.02 for the spectral index. Although the power-law fit cannot be rejected, from Figure \ref{Fig.3} one can see a noticeable steeping of an initially hard ($E^{-2}$ type) spectrum above $\approx 50$~TeV. A hard spectrum at low energies is demanded also by the flux upper limit set by Fermi LAT around 10GeV (see below). {\bf Therefore we fit the spectrum also by a log-parabola function  $f(E)=J\cdot({E}/{E_{0}})^{-(\alpha+\beta log_{10}({E}/{E_{0}}))}$. The result is shown in Figure \ref{Fig.3}. The obtained best-fit parameters are $~\rm J=(3.7\pm0.6)\times10^{-16}~\rm TeV^{-1}~\rm cm^{-2}~\rm s^{-1}$,$~\rm \alpha=2.4\pm0.4$ and $~\rm \beta=2.6\pm1.3$, with $\chi^2/ndf=0.26/2$. The improvement is about 2.5$\sigma$ comparing to a single power-law fit, so we can not get a solid conclusion now. With the accumulation of data, the behavior of the spectrum will be more clear.}

\section{Discussions}
\subsection{Multiwavelength observations}

4FGL J0340.4+5302 is the closest GeV $\gamma$-ray source in the latest {\sl Fermi} LAT source catalog (4FGL-DR2) \citep{4fgl} at an angular distance of $0.16^{\circ}$, which is within the extension of LHAASO J0341+5258. This source is labeled as an unidentified source in the catalog. To investigate the relationship between 4FGL J0340.4+5302 and LHAASO J0341+5258,  we analyzed 12 years of {\sl Fermi}  Pass 8 data (from August 4, 2008 until  August 14, 2020), and performed binned likelihood analysis in the $20^{\circ}\times20^{\circ}$ region centered on LHAASO J0341+5258 for \grays in the energy range of 0.1--500~GeV. We found that using the Gaussian disk model for 4FGL J0340.4+5302, the 95\% C.L. upper limit of the extension ($\sigma_{\rm disk}$) is $\le0.3^\circ$  while the TS value of the extension hypothesis is $\sim 0.15$. By comparing the LogParabola model to power-law model, we find its spectrum is significantly curved. 
The replacement of the point source model with a Gaussian disk of $\sigma_{disk}=0.29^{\circ}$  located at the center of LHAASO J0341+5258 does not influence the results. To compare with LHAASO measurement, in Fig.\ref{fig:sed} we present the SEDs and upper limits using the same spatial template of LHAASO J0341+5258.

In the X-ray domain, there are 4 sources in the second ROSAT all-sky survey source catalogue (0.1$-$2.4 keV) within $0.6^\circ$  from the center of LHAASO J2031+5258. The angular distance for the 4 X-ray sources, named as 2RXS J034125.8+525530, 2RXS J033928.5+530720, 2RXS J034316.5+524331 and 2RXS J034203.0+532329 \citep{2016A&A...588A.103B}, are $0.046^\circ$, $0.321^\circ$, $0.379^\circ$ and $0.434^\circ$ separately. {\bf  The flux of all these 2RXS sources are not well measured, the positions of 2RXS J033928.5+530720  and 2RXS J034316.5+524331 are coincident with 2SXPS 172133 and 2SXPS 171354 in the 2RXPS Swift X-ray telescope point source catalogue \citep{2020yCat.9058....0E}, respectively. The energy flux of 2SXPS 171354 and 2SXPS 172133 are $(1.6 \pm 0.7) \times 10^{-13}\, \rm erg\,cm^{-2}\,s^{-1}$ and $(5.5 \pm 1.6) \times 10^{-13}\, \rm erg\,cm^{-2}\,s^{-1}$ from the catalog, respectively. }

We note that the LHAASO J0341+5258 is indeed extended and the flux of X-ray point sources may come from a much compact region and can have different origins. In this regard, we found that the archived Chandra ACIS observation with ID 16828 is partially overlapped  with LHAASO J0341+5258. No significant emission was detected in this region. The upper limit was derived by assuming the observed X-ray flux in the overlapped region is dominated by Poisson noises. We  summed over the overlapped region  to get the  total photon counts N, and used $3\sqrt{N}$ as the 3$\sigma$ upper limits of the total counts from the source. Then we divided the counts by the exposure to derive the upper limits of flux.  Since the FoV of Chandra is significantly smaller than the derived intrinsic extension of LHAASO J0341+5258, the upper limit is then scaled by the area. Finally, the derived upper limit for LHAASO J0341+5258 is $1.5\times 10^{-12}\, \rm erg\,cm^{-2}\,s^{-1}$ in the energy range of 0.2$-$8.8 keV. It should be noted that we neglect the absorption in the estimations, but the effects should be minor in the whole energy range, unless the column density in the line of sight is extremely large and the intrinsic X-ray flux is very soft. 

{\bf HAWC detected no significant source in this region \citep{2020ApJ...905...76A}, but with the online interactive tools \footnote{https://data.hawc-observatory.org/datasets/3hwc-survey/coordinate.php} the derived $2 \sigma$ flux upper limit  is $1.1 \times 10^{-12}~\rm erg~cm^{-1}~s^{-1}$,  
assuming a spectrum index of -2.5 and extension of 0.5$^{\circ}$ with a reference energy of 7 TeV. }
 

\begin{figure}
\centering
\plotone{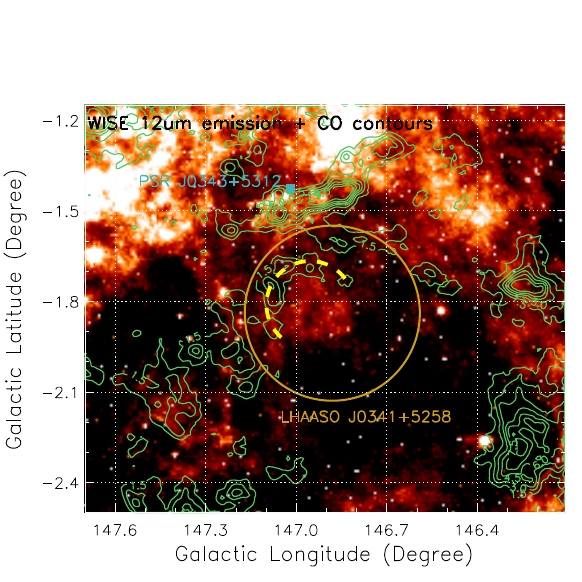}
\caption{WISE 12$\mu$m emission and CO emission (green contours) toward
LHAASO J0341+5258 (the golden cycle). The CO emission is integrated
in the velocity interval of 0--10~km~s$^{-1}$ and starts from
1.5~K~km~s$^{-1}$ with a step of 1.5~K~km~s$^{-1}$. \textbf{The yellow-dashed line marks the half-shell CO structure.} The cyan box indicates PSR J0343+5312.}
\label{Fig.4}
\end{figure}

The CO observations of this region from Milky Way Imaging Scroll Painting project\citep{2019ApJS..240....9S} is shown in Fig \ref{Fig.4}. \textbf{The molecular gas is partly overlapped with LHAASO J0341+5257, which presents a half-shell structure, labeled as yellow-dashed line.} The total mass of gas within $1^{\circ}$ of the source is about $10^3 ~M_{\odot}$ considering a distance of 1 kpc. Assuming the average cloud thickness of $0.5^{\circ}$, the $\rm H_2$ cubic density is about 50~$\rm cm^{-3}$. There is no clear CO emission at larger distances. \textbf{ Meanwhile, although the total CO emission is not bright in this region, $^{13}$CO line is also detected at the enhanced $^{12}$CO emission region of the half-shell structure, which implies the existence of dense clumps. The critical density of $^{13}$CO is about $2000$ cm$^{-3}$. So the mean density of several tens cm$^{-3}$ is reasonable for the revealed MC in the field of view of the LHAASO source. Also in Fig \ref{Fig.4} we can see a significant 12$\mu$m infrared radiation surrounded by the half-shell structure, which is inside the TeV source.} One may argue for a hint that the infrared emission, molecular gas and TeV emission are physical associated with each other, but further observations are needed to draw a firm conclusion. 




\begin{figure}
\centering
\includegraphics[width=0.8\linewidth]{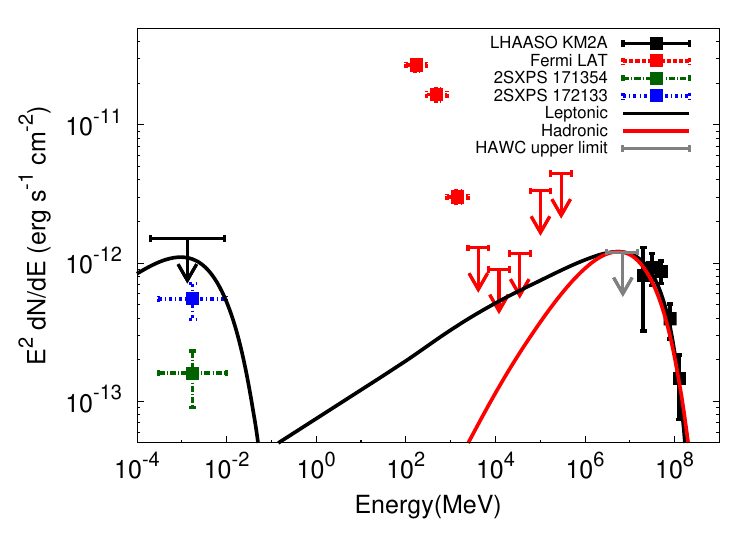}
\caption{\bf The multiwavelength SEDs of LHAASO J0341+5258. The black squares are the LHAASO KM2A observations, the red points and arrows are the Fermi LAT spectral points and upper limits, the green and blue point is the X-ray flux for 2SXPS 171354 and 2SXPS 172133 from the 2RXPS catalog \citep{2020yCat.9058....0E}, respectively.  The black arrow labels the upper limit derived from Chandra observations. The gray arrow is the HAWC upper limit from the online interactive tools (see the text for detail). Also shown is the phenomenological fitting in both leptonic (black curves) and hadronic scenarios (red curves). }
\label{fig:sed}

\end{figure}

\subsection{Phenomenological fitting and possible origin}


Due to the spatial coincidence between LHAASO J0341+5258
and 4FGL J0340.4+5302, we consider at first the possibility that both sources have a unified origin. However, we note that the energy flux of \fermi LAT observations extend to more than 1 GeV and is one order of magnitude higher than  TeV \grays, which makes the synchrotron origin of the MeV-GeV emission quite unlikely. On the other hand, the spectral shape of 4FGL J0340.4+5302 reveals a sharp cutoff feature, which is quite similar to GeV \gray emission of pulsars.  If this is true, the TeV emission can be interpreted as originating from the inverse Compton scatterings (ICs) of the relativistic electrons in the surrounding pulsar wind nebular (PWN) or or a larger structure enveloping the PWN \citep{aharonian04}, the so called pulsar halo \citep{sudoh19}. The observed TeV spectrum is steep, which implies that we probably observed the cutoff regions. Thus, for simplicity, we used a power-law electron spectrum with a superexponential cutoff $f(E) \sim E^{-\Gamma} exp(-(E/E_c)^2)$. Since the KM2A data points are mainly at the cutoff region, the constraining to the index $\Gamma$ is loose. We fix $\Gamma$ to 2.5 to provide a schematic fit to the data. Taking into account the CMB and interstellar radiation fields (ISRFs) in the solar vicinity \citep{popescu17} as the target photon fields, the fit is labeled as black lines in Fig.\ref{fig:sed}.  The derived $E_c$ is 200 TeV, and the total electron energy  above 10 TeV is about $10^{44} ~ \rm erg (\frac{d}{1~\rm kpc})^2$, where d is the distance to the source. The X-ray upper limit can also be used to set the upper limit for the magnetic fields, which is $3~\rm \mu G$.  We note that the X-ray upper limit was derived from a much smaller region than the VHE emission region, a dedicated study with the instruments with larger FoV is required to get more robust constraints.  The upper limit of $3~\rm \mu G$ is feasible for both the PWN or pulsar halo case. Due to the unknown distance and thus the unknown physical size of this source, it is hard to distinguish the PWN or pulsar halo scenarios.  The search for pulsation of the GeV data of 4FGL 0340.4+5302 would be useful to test such a scenario.  And Version 1.21 of the ATNF (Australia Telescope National Facility) Pulsar Catalogue lists another pulsar within a angular distance of $0.4^{\circ}$ away from LHAASO J0341+5258 \citep{1978ApJ...225L..31D}. 
PSR J0343+5312 has a distance of 2.48 kpc, age of 2.28 Myr and spin-down luminosity of $7.3\times10^{31} \rm erg/s$, which seems too weak and too old to account for the VHE $\gamma$-ray emissions, even in the pulsar halo scenario.

Finally, although there is no known hadronic accelerators such as SNRs in this region, we cannot formally rule out the hadronic origin. Indeed as mentioned in last section, there are hints that the TeV emissions are correlated with infrared emission as well as a weak CO shell structure. The ambient gas density of $50~\rm cm^{-3}$ is used in the phenomenological hadronic fittings. To fit the LHAASO KM2A data points, we found that for parent protons with spectral index  larger than 2, the predicted \gray spectrum will violate the upper limit at 10~GeV derived from Fermi LAT data (see Figure.\ref{fig:sed}).   We present a schematic fit to the \gray data by fixing the index to 1.5 and  with the best-fit proton spectrum of $f(E) \sim E^{-1.5} exp(-(E/190~\rm TeV))$,  which is labeled as red curve in Fig.\ref{fig:sed}.  Note that this spectrum is too hard compared to the cosmic ray  spectrum   predicted by the Diffusive Shock Acceleration applied to young SNRs.  On other hand, it can be naturally explained by cosmic rays that have escaped the SNR  and hit a cloud located within  $\sim~100$ pc proximity of  a middle-aged or older (presently invisible) SNR \citep{gabici07}. The derived total energy in protons above 10 TeV is {\bf $1.5\times 10^{46}  (\frac{d}{1~\rm kpc})^2(\frac{50~\rm cm^{-3}}{n})~ \rm erg$ }, where n is the ambient gas density.

\section{Conclusions}
Since December 2019, the half of LHAASO-KM2A experiment has monitored  the  sky in the declination band from -15$^{\circ}$  to 75$^{\circ}$ above tens of TeV with high duty cycle. An excess with a pre-trial significance of $8.2\sigma$ was detected from the direction of LHAASO J0341+5258 using events with energy above 25 TeV. LHAASO J0341+5258 is an extended source with a width of 
$\sigma = (0.29 \pm 0.06_{stat}\pm0.02_{sys})^{\circ}$. 
The energy spectrum can be approximately described by a power-law, even though there is hint of curvature at around 50 TeV.  The integrated energy flux of $\gamma$-ray emissions above 25 TeV is $1.44\times10^{-14}($cm$^{-2}$s$^{-1})$,  which accounts for about $20\%$ of the flux from the Crab Nebula.

The source is positionally coincident with a  GeV $\gamma$-ray source 4FGL J0340.4+5302 observed by {\sl Fermi} LAT \citep{4fgl}. 
The upper limit of the latter at 10 GeV demands a hard spectrum of LHAASO J0341+5258 (photon index $<$ 2) when extrapolating toward low energies.  Together with the fast drop of the $\gamma$-ray fluxes above 50 TeV, the phenomenological fitting in the leptonic and hadronic scenarios the power-law require the indices of the parent electrons and protons  to be 2.5 and 1.5,  with cutoffs  of about 200 TeV and 190 TeV, respectively.  The most likely  realization of the leptonic scenario is the extended emission of a PWN and/or a Pulsar Halo. The challenge of this scenario is the lack of a reported powerful pulsar. Interestingly, such a pulsar could be the gamma-ray source 4FGL J0340.4+5302 with a characteristic spectrum below 1 GeV. The detection of pulsed radio emission from this source would support the IC origin of the UHE gamma-ray emission. The hadronic origin of the UHE emission can be interpreted as an “echo"  from a molecular cloud(s) $-$ the result of interactions of protons  with dense gas regions in the proximity of  an old (currently invisible) SNR.


This work is supported in
China by National Key R\&D program of China under the grants 2018YFA0404201, 2018YFA0404202,  2018YFA0404203, 2018YFA0404204, by NSFC (No.12022502, No.11905227, No.11635011,  No.U1931112),  and  in Thailand by RTA6280002 from Thailand Science Research and Innovation. The authors would like to thank all staff members who work at the LHAASO site above 4400 meters above   sea level year-round to maintain the detector and keep the electrical power supply and other components of the experiment operating smoothly. We are grateful to the Chengdu Management Committee of Tianfu New Area for their constant financial support  of research  with LHAASO data.

This research made use of the data from the Milky Way Imaging Scroll Painting (MWISP) project, which is a multi-line survey in $^{12}$CO/$^{13}$CO/$^{18}$CO along the northern galactic plane with PMO-13.7m telescope. We are grateful to all the members of the MWISP working group, particularlly the staff members at PMO-13.7m telescope, for their long-term support. MWISP was sponsored by National Key R\&D Program of China with grant 2017YFA0402701 and  CAS Key Research Program of Frontier Sciences with grant QYZDJ-SSW-SLH047.

\appendix
\section{Fermi-LAT analysis}

Since LHAASO J0341+5258 is extended, we firstly use {\sl fermipy} \citep{Wood2017} to reinvestigate the location and extension of 4FGL J0340.4+5302 using events with energy range of 1$-$500 GeV and the Gaussian disk model. We found that the best-fit positions of both localization and extension test is basic consistent with the original location of 4FGL J0340.4+5302 in the 4FGL catalog, moreover, the TS value of extension is only $\sim0.15$ and the 95\% C.L. upper limit of the extension ($\sigma_{\rm disk}$) is $\le0.3^\circ$, which is consistent to the results of KM2A observation. However, its central position is about $0.15^\circ$ away from LHAASO J0341+5258. 
To test whether the spectrum of 4FGL J0340.4+5302 is curved,  we compared  a simple power-law model 
to a LogParabola model, i.e., $\mathrm{d}N/\mathrm{d}E = N_0 (E/E_\mathrm{b})^{-\alpha - \beta \log(E/E_\mathrm{b})}$.
The likelihood of the LogParabola model $\mathcal{L}_{\rm LogP}$ is increased by $\sim 125$ compared to that of the power-law model $\mathcal{L}_{\rm LogP}$. Thus $TS_{\rm curve}\equiv 2{\rm log} (\mathcal{L}_{\rm LogP} /\mathcal{L}_{\rm PL})= 250$,  which means that the spectrum of 4FGL J0340.4+5302 is  significantly curved. 
For 4FGL J0340.4+5302, its best-fit spectral parameter of LogParabola  model $\alpha=3.15\pm0.05$, $\beta=0.51\pm0.05$, 
and the  energy flux is $\sim 5.0\times 10^{-11} \rm erg\,cm^{-2}\,s^{-1}$ 
in the energy range of 0.1-500\,GeV, corresponding to a significance of $\sim 56\sigma$ (TS $\simeq 3149$).

Next, we replaced the spatial model of  4FGL J0340.4+5302 with a gaussian disk of $\sigma_{\rm disk}=0.29^{\circ}$ located at the center of LHAASO J0341+5258, as an representative of LHAASO J0341+5258 at GeV band. Through the likelihood analysis, we found that the spectral shape and total flux of this GeV LHAASO source is very similar to those of 4FGL J0340.4+5302 as we obtained above, but such replacement does not increase the total likelihood compared to the original \fermi  4FGL-DR2 source model. 


Then we divided the energy range 0.1$-$500\,GeV into eight logarithmically spaced energy bins and  derived the SEDs using the 4FGL catalog point-like source J0340.4+5302  and the GeV representative of LHAASO J0341+5258, respectively. We found in both cases, the overall spectra are very soft and no significant emissions above 2 GeV is detected, thus only upper limits are derived for their flux above 2 GeV. 



\bibliography{sample63}{}
\bibliographystyle{aasjournal}



\end{document}